\documentclass[12pt]{article}
\usepackage{amssymb}\usepackage{bm}\usepackage{amsfonts}
\newcommand{\beq}{\begin{equation}}\newcommand{\eeq}{\end{equation}}\newcommand{\beqa}{\begin{eqnarray}}
\newcommand{\eeqa}{\end{eqnarray}}\newcommand{\ts}{\textstyle}
\newcommand{\dl}{\bm{\delta}}\newcommand{\nn}{\nonumber}\newcommand{\vep}{\varepsilon}

\begin{document}
{\renewcommand{\thefootnote}{\fnsymbol{footnote}}
\hfill  IGC--08/2--2\\
\medskip
\begin{center}
{\LARGE  Canonical Lagrangian Dynamics and General Relativity}\\
\vspace{1.5em}
Andrew Randono\footnote{e-mail address: {\tt arandono@gravity.psu.edu}}
\\
\vspace{0.5em}
Institute for Gravitation and the Cosmos,\\
The Pennsylvania State
University,\\
104 Davey Lab, University Park, PA 16802, USA\\
\vspace{1.5em}
\end{center}
}

\setcounter{footnote}{0}

\abstract{Building towards a more covariant approach to canonical classical and quantum gravity we outline an approach
to constrained dynamics that de-emphasizes the role of the Hamiltonian phase space and
highlights the role of the Lagrangian phase space. We identify a ``Lagrangian one-form" to replace the standard
symplectic one-form, which we use to construct the canonical constraints and an associated constraint algebra. The
method is particularly useful for generally covariant systems and systems with a degenerate canonical symplectic form,
such as Einstein Cartan gravity, to which we apply it explicitly. We find that one can compute the constraint algebra and demonstrate the closure
of the constraints without gauge fixing the Lorentz group or introducing primary constraints on the phase space
variables. Ultimately our aim is towards a more covariant approach for canonical quantum gravity, and we discuss a possible route to quantization. Applying these techniques and using methods from geometric quantization, we find a new representation of the pre-quantum operator corresponding to the Hamiltonian constraint, which in contrast to the standard representation, is a kinematical operator that simply generates timelike diffeomorphisms on functionals of the Lagrangian phase space . This opens the possibility for a kinematical {\it spacetime}--diffeomorphism invariant Hilbert space.}

\section{Motivation}
As a first step in developing a more covariant framework for the Hamiltonian dynamics of generally covariant systems, 
here we outline a new perspective on symplectic geometry and its relation to the Lagrangian dynamics. The framework we
will develop is applicable to many different systems, but it is especially relevant to diffeomorphism
invariant systems. The method can be seen as a generalization of Hamiltonian mechanics for systems where the 
symplectic form is degenerate. In particular, we
have tailored the formalism to suit general relativity in the first-order Einstein Cartan formalism, where the natural
symplectic form on the unreduced phase space has this property. Our main goal is to construct a formalism for canonical
Einstein-Cartan gravity where the full local Lorentz group is retained explicitly and one does not have to introduce
primary constraints on the dynamical variables. We will begin such a construction here, but an explicit calculation of the constraint algebra and discussion of the interesting geometric features that emerge for a follow-up paper \cite{Randono:CovariantCanonical}.

As we will see, the (pre)symplectic form for Einstein-Cartan gravity is degenerate, even prior to any restriction to a constraint submanifold. There are several standard approaches to
dealing with this degeneracy. The most obvious would be to factor out the degenerate gauge orbits of the symplectic
form prior to implementing Hamilton's equations. Though this may be appropriate for many physical systems, due to the
complexity of general relativity, this procedure does not appear to be tractable in practice. Thus, most approaches
resort to the alternative tactic of gauge fixing the Lorentz symmetry of the system in order to arrive at a degenerate
symplectic form. Loop quantum gravity relies on such an approach where the gauge symmetry is reduced from $Spin(3,1)$
to $SU(2)$ by partially fixing to the time gauge (see e.g. \cite{Ashtekar:book,Ashtekar:LQGReview1}). A third approach,
often called covariant canonical gravity \cite{Alexandrov:Covariant,Livine:Covariant},
does not gauge fix, but rather enlarges the phase space such that the symplectic structure on the larger phase space is non-degenerate. To regain general relativity one must then introduce primary constraints on the momenta that reduce the phase space back down to the original phase space on which the now pre-symplectic form is degenerate. Unfortunately, these primary constraints and the standard Hamiltonian constraints do not close under the action of the Poisson, so one must introduce new second class constraints on the dynamical variables.
In this work we draw attention to the root of the problem: in the case of a degenerate symplectic form,
\textit{Hamilton's equations themselves} define constraints on the phase space, and when this is taken into account, a
self-consistent canonical framework emerges which does not involve factoring out gauge orbits, fixing the gauge, or
introducing primary class constraints on the phase space variables.

The problems stemming from a degenerate symplectic form are two-fold:
\begin{itemize}
\item{First, the degeneracy implies that there are vector fields $\bm{\bar{Z}}$ for which 
\beq
\bm{\Omega}(\bm{\bar{Z}},\ )=0.
\eeq
Thus, suppose we had a function, $f$, that can be related to a Hamiltonian vector $\bm{\bar{X}}_{f}$ by the usual
relation $\bm{\Omega}(\bm{\bar{X}}_{f},\ )=\dl f$, where $\dl$ is the exterior derivative on the Hamiltonian phase
space. Then the vector field $\bm{\bar{X'}}_{f}=\bm{\bar{X}}+\bm{\bar{Z}}$ would also be a canonical vector field to
$f$ since $\bm{\Omega}(\bm{\bar{X'}}_{f},\ )=\dl f$. Thus, canonical vector fields associated to a functional are not
unique.
}
\item{Second, a potentially more serious problem, there exist functionals $f$, which cannot be associated with any
Hamiltonian vector field in the usual way. To see this, suppose we had a functional $f$ such that $\bm{\Omega}(\bm{\bar{X}}_{f},\
)=\dl f$ everywhere. Then, inserting a vector $\bm{\bar{Z}}$ that is in the kernel of $\bm{\Omega}$, we see that we must
have\footnote{We will employ two notations for the insertion of a vector field $\bar{V}$ in the first slot of a differential form $\alpha$: $\iota_{\bar{V}}\alpha = \alpha(\bar{V})\equiv \alpha(\bar{V},\ ,\,...\,,\ )$.}
\beq
\bm{\Omega}(\bm{\bar{X}}_{f},\bm{\bar{Z}})=\iota_{\bm{\bar{Z}}}\dl f=\mathcal{L}_{\bm{\bar{Z}}}f=0. 
\eeq
Thus, $f$ must be constant along the integral lines of $\bm{\bar{Z}}$, which is obviously not true for every
functional.
}
\end{itemize}
From the above problems, we conclude that we need to rethink our notion of a Poisson bracket in context of a
degenerate symplectic form. We will show that one can make sense of the Poisson bracket of constraints precisely
when Hamilton's equations of motion hold.

\section{Covariant symplectic structures on the solution submanifold \label{StandardCovariant}}
In this section we will review the standard formulation of covariant symplectic dynamics. Our main goal is to highlight the main differences between the these approaches and the framework we will develop in this paper. 

In reaction to the lack of explicit covariance of the standard Hamiltonian framework, many attempts at making the Hamiltonian framework more covariant have been proposed in the past, and the subject has a long history \cite{Dedecker, Kijowski1, Kijowski2, Garcia:Symplectic, AshtekarMagnon, AABombelli:CovariantSymplectic, Crnkovic, CrnkovicWitten, Aldaya:CovariantCanonical, Christodoulou}. Many of these attempts rely on a reduction of the Lagrangian phase space to critical sections of the variation principle, the submanifold of the space of histories formed by solutions to the equations of motion, and it is from these approaches that we would like to discriminate. This discussion will follow closely along the lines of \cite{AshtekarMagnon, AABombelli:CovariantSymplectic, Crnkovic, CrnkovicWitten, Aldaya:CovariantCanonical}.

Typically it is assumed that the n-dimensional base manifold is topologically $M=\mathbb{R}\times \Sigma$, where $\Sigma$ is topologically a three-manifold. We will assume the manifold has two spacelike hypersufaces $\Sigma_1$ and $\Sigma_2$ as endcaps located at $t=t_1$ and $t=t_2$ so that $\partial M=\Sigma_1\cup \Sigma_2$ (one can also include the boundary at spatial infinity, but for simplicity we will ignore this boundary). Consider the action of the generic form
\beq
S=\int_M \widetilde{L}
\eeq 
where $\widetilde{L}=\widetilde{L}(\Phi^{(i)}, d\Phi^{(j)})$ is the Lagrangian n-form of the dynamical fields $\Phi^{(i)}$ and their derivatives $d\Phi^{(j)}$. Consider an arbitrary variation of the dynamical fields denoted $\dl$, which we interpret as the exterior derivative on the Lagrangian phase space, or space of histories denoted $\Gamma_L$. Typically an arbitrary variation of the action splits into a bulk term, and a boundary term:
\beq
\dl S=\int_M \widetilde{\bm{\theta}}+\int_{\partial M} \widetilde{\bm{J}},
\eeq
or, written as a pointwise variation, we have
\beq
\dl \widetilde{L}=\widetilde{\bm{\theta}}+d\widetilde{\bm{J}}\ .\label{PointwiseVariation}
\eeq
Here we have employed a notation that we will use throughout this paper where we denote vectors and forms over the infinite dimensional Lagrangian phase space in $\bm{bold}$. The tilde over integrands indicates that the corresponding object is also a differential form on the base manifold. 

The bulk Lagrange equations of motion are then equivalent to $\widetilde{\bm{\theta}}=0$. The standard covariant approach then proceeds by restricting attention to the submanifold $\Gamma^{(0)}_L$ of the Lagrangian phase space consisting only of solutions to equations of motion. In addition, in this construction, one must restrict not only the points of $\Gamma_L$, but also the variation $\dl$ to only those variations that preserve the equations of motion. With these restrictions, the ``symplectic current" is defined as
\beq
\widetilde{\bm{\omega}}\equiv  \dl \widetilde{\bm{J}}\,\Big|_{\Gamma^{(0)}_L}\ .
\eeq
The key identity that follows from (\ref{PointwiseVariation}) by noting $\dl\dl \equiv 0$ is the conservation equation of the symplectic current when restricted to the solution submanifold:
\beq
d\widetilde{\bm{\omega}}=0\,.
\eeq
Integrating the above over all of $M$ we have
\beq
\int_M d\widetilde{\bm{\omega}}=\int_{\Sigma_2}\widetilde{\bm{\omega}}-\int_{\Sigma_1}\widetilde{\bm{\omega}}=0
\eeq
from which we define the pre-symplectic form on any hypersurface
\beq
\bm{\Omega}\equiv \int_{\Sigma_1}\widetilde{\bm{\omega}}=\int_{\Sigma_2}\widetilde{\bm{\omega}}\,.
\eeq
Since $\Sigma_1$ and $\Sigma_2$ are arbitrary spacelike hypersurfaces, the pre-symplectic form defined as such is independent of the hypersurface on which it is defined. This two-form on $T^* \Gamma^{(0)}_L$, is highly degenerate, hence the name ``pre-symplectic". This is due to the restriction to the solution submanifold: the degenerate directions formed by vector fields, $\bm{\bar{Z}}$, such that $\bm{{\Omega}(\bar{Z},\ )}=0$ generate gauge transformations that map solutions to the equations of motion to gauge equivalent solutions. 

For future reference it will be useful to write the hypersurface independence of the pre-symplectic form in a different way. Suppose $M$ is a ``thin--sandwich" embedded in a larger manifold $M^*$ so that $\Sigma_2$ is can be constructed from $\Sigma_1$ by an infinitesimal timelike diffeomorphism generated by a timelike vector field $\bar{t}$, denoted $\phi_{\bar{t}}(\Sigma_1)=\Sigma_2$. From the conservation equation we then have
\beqa
\int_{\Sigma_1}\widetilde{\bm{\omega}}&=&\int_{\Sigma_2=\phi_{\bar{t}}(\Sigma_1)}\widetilde{\bm{\omega}}\nn \\
&=&\int_{\Sigma_1}\widetilde{\bm{\omega}}+\mathcal{L}_{\bar{t}}\,\widetilde{\bm{\omega}}\,.
\eeqa
where $\mathcal{L}_{\bar{t}}=\iota_{\bar{t}}\,d+d\,\iota_{\bar{t}}$ is the Lie derivative on $M$. Thus, the conservation equation can be written $\int_{\Sigma}\mathcal{L}_{\bar{t}}\,\widetilde{\bm{\omega}}=0$. Defining the time evolution vector field on $T\Gamma_L$ by $\bar{\bm{t}}=\int_M \mathcal{L}_{\bar{t}}\Phi^{(i)} \frac{\dl}{\dl \Phi^{(i)}}+\mathcal{L}_{\bar{t}}d\Phi^{(j)} \frac{\dl}{\dl d\Phi^{(j)}}$ we can rewrite the conservation equation as
\beq
\bm{\mathcal{L}_{\bar{t}}\,\Omega} =\int_{\Sigma}\mathcal{L}_{\bar{t}}\,\widetilde{\bm{\omega}}=0\ . \label{SymplecticConservation}
\eeq
where $\bm{\mathcal{L}_{\bar{t}}}=\bm{\iota_{\bar{t}}}\,\dl+\dl \,\bm{\iota_{\bar{t}}}$ is the Lie derivative on $\Gamma_L$. This formula will be useful later for the purposes of comparison.

Though we will use some of the ideas introduced in these covariant approaches, we will not proceed along this route. In particular, we wish to avoid any restriction to a solution submanifold. Our reason for avoiding this restriction are inspired by quantum theory. Eventually we wish to develop a formalism that is more appropriate for carrying covariant techniques to the quantum theory, and we will briefly discuss one route to doing this towards the end of this paper. With an aim towards quantum theory, this approach is not particularly useful for two reasons. First, the solution submanifold is often difficult to characterize in practice, particularly with regards to diffeomorphism invariant theories like general relativity. Most considerable progress in canonical quantization of constrained systems fall into the category of ``quantize then constrain" as opposed to ``constrain then quantize" approaches (with the possible exception of topological field theories where the critical points of the variational principle lie in isolated regions of the phase space separated by topological characteristics). A quantization scheme employing the above scenario prior to quantization would be an extreme example of the latter. Secondly, with regards to systems that have local degrees of freedom, as clearly explained in the conclusion of \cite{AABombelli:CovariantSymplectic}, such a quantization would not be very interesting. Specifically, since the covariant phase space $\Gamma^{(0)}_L$ consists entirely of classical solutions, such a quantization would have no avenue for genuine, pure quantum mechanical effects resulting from off-shell contributions which play a role in, for example, quantum tunneling effects, and are believed to be responsible for singularity resolution in quantum gravity. Thus, we wish to develop covariant Hamiltonian techniques that avoid any restriction to classical submanifolds of the phase space, but we also need to retain the aspects of Hamiltonian dynamics that make it amenable to quantization.
 
\section{Classical Hamiltonian dynamics}
The framework we develop in this paper will be sufficiently unfamiliar that it will be helpful to first discuss a
simple familiar model. As we will see, the ideal setting for this framework is a generally-covariant system, but a
simpler model will help to introduce the basic ideas. The setup is similar to that above: we consider the bulk and boundary terms that emerge from the variational principle. However, in this case, and especially its extension to diffeomorphism invariant theories,  it will be more useful to work directly with the action as opposed to the Lagrangian. In addition, it will convenient to work with a single timelike boundary, sending the initial boundary to past infinity, where we will assume the arbitrary variation vanishes. This is simply a matter of convenience, and is in keeping with the general philosophy that the Hamiltonian approach is simply a set of rules for advancing a spacelike hypersurface in time given all past initial data.

With this in mind, we begin by considering the action of a non-relativistic particle in a potential $V(x)$. The action, is given by
\beq
S=\int^t_{-\infty}\left(\ts{\frac{1}{2}}m\,\dot{x}\cdot \dot{x}-V(x)\right)dt\,.
\eeq
Consider now an arbitrary variation of the action given by $\dl S$. As before, we can think of $\dl$ as the exterior derivative
on the space of histories or Lagrangian phase space. The variation has a bulk piece and a boundary piece:
\beqa
\dl S &=& \int^t_{-\infty} \left(m\dot{x}\cdot \frac{d}{dt}\dl x-\nabla V\cdot \dl x\right)dt \\
&=& m\dot{x}\cdot \dl x\Big|_t-\int^t_{-\infty}\left(m\ddot{x}+\nabla V\right)\cdot \dl x\, dt\,.
\eeqa
We define the symplectic one form or symplectic potential, $\bm{J}$, to be the boundary piece of the variation:
\beq
\bm{J}\equiv m\dot{x}\cdot \dl x\Big|_t\,.
\eeq
The bulk piece, which we will refer to as the Lagrangian one-form, $\bm{\theta}$, is given by:
\beq
\bm{\theta}\equiv -\int^t_{-\infty}\left(m\ddot{x}+\nabla V\right)\cdot \dl x\, dt\,.\label{Lagrangian1form}
\eeq
As previously, the negative of the exterior derivative of the symplectic one-form gives the symplectic form:
\beq
\bm{\Omega}=-\dl\bm{J}=\dl x \wedge \dl(m\dot{x})\Big|_t\,.
\eeq
However, since we are not restricting functionals and variations to the solution submanifold where $\bm{\theta}=0$, we have additional freedom.  In particular, since $\dl$ is the exterior derivative, from the general identity $\dl\dl=0$ we have
 \beq
 \dl\dl S =\dl\bm{J}+\dl\bm{\theta}=0\,.
 \eeq
 Thus, we see that $\dl\bm{\theta}=-\dl\bm{J}=\bm{\Omega}$, and this can be verified explicitly from
(\ref{Lagrangian1form}).
Because of this identity, and for other reasons that we will see shortly, it will be advantageous to focus our
attention on the Lagrangian one-form as opposed to the symplectic one-form. 

It is important to note at this point all of the objects, $\bm{\theta}$, $\bm{J}$, and $\bm{\Omega}$ depend explicitly on the time variable, t, or more generally, the choice of initial hypersurface. This is in contrast to the covariant phase space formalism detailed above, where the symplectic form was hypersurface--independent by construction, a property that follows from the restriction to the solution submanifold of the Lagrangian phase space. In the current context, the time independence of the symplectic form is implemented directly by Hamilton's equations of motion, which state that the time evolution of the initial data must be a symplectomorphism.

Consider the time translation vector field $\bm{\bar{t}}$ given by:
\beq
\bm{\bar{t}}=\dot{x}\cdot\frac{\dl}{\dl x}+\ddot{x}\cdot \frac{\dl}{\dl \dot{x}}\,.
\eeq
The change in the action under a time translation is a pure boundary term. Explicitly it is given by
\beqa
\bm{\theta(\bar{t})}&=&\int^t_{-\infty} -(m\ddot{x}+\nabla V)\cdot\dot{x}\,dt \nn\\
&=& \int^t_{-\infty} -(m\ddot{x}+\nabla V)\cdot dx\,.
\eeqa
The work energy theorem then tells us that the above integral is the negative of the total mechanical energy, or the
Hamiltonian (up to a constant that is fixed by the boundary conditions at past infinity):
\beq
-\bm{\theta(\bar{t})}=\left(\ts{\frac{1}{2}}m v^{2}+V(x)\right)\Big|^t_{-\infty}=H-E_{-\infty}\,.
\eeq

We can now write Hamilton's equations in a non-standard way. Consider the Lie derivative of the Lagrangian one-form:
\beq
\bm{\mathcal{L}_{\bar{t}}\,\bm{\theta}}=\bm{\iota_{\bar{t}}\,\dl\bm{\theta}}+\bm{\dl \,\iota_{\bar{t}}\theta}\,.
\eeq
From the identities $\dl\bm{\theta}=\bm{\Omega}$ and $-\bm{\theta(\bar{t})}=H-E_{-\infty}$, we have
\beq
\bm{\mathcal{L}_{\bar{t}}}\,\bm{\theta}=0 \ \ \longleftrightarrow \ \ \bm{\Omega(\bar{t},\ )}=\dl H\,.
\eeq

We recognize the last equation on the right as Hamilton's equations. Thus, we see that Hamilton's equations are equivalent to the time invariance of the Lagrangian one-form. Likewise, they imply the symplectic form is time invariant. In general, given a function, $f$, and its associated canonical vector field, $\bm{\bar{X}}_f$, defined such that $\bm{\Omega(\bar{X}_f,\ )}=\dl f$, it follows that $\bm{\mathcal{L}_{\bar{X}_f}\Omega}=0$, and we say that $\bm{\bar{X}}_f$ is the generator of a symplectomorphism. The non-trivial part of Hamilton's equations is the identification of canonical vector field paired to the Hamiltonian with the time--evolution vector field, $\bm{\bar{t}}\,$. Thus, we have
\beq
\bm{\mathcal{L}_{\bar{t}}\,\Omega}=0\,.
\eeq 
Comparing this with (\ref{SymplecticConservation}) we see that hypersurface-invariance is regained by the implementation of Hamilton's equations. 

In contrast, the symplectic one-form is not time invariant. However, from Hamilton's equations it transforms by the addition of an exact form under time evolution:
\beqa
\bm{J} &\rightarrow& \bm{J}+\bm{\mathcal{L}_{\bar{t}}J} \nn\\
&=& \bm{J}+\dl(\bm{J(\bar{t})}-H)\,.
\eeqa
In geometric quantization, which we will discuss at the end of this paper, $\bm{J}$ is interpreted as a $U(1)$ connection and the above transformation is identified with a $U(1)$-transformation. 

\subsection{Extension to diffeomorphism invariant field theories}
We now wish to extend this procedure to diffeomorphism invariant field theories. Recall, our main goal is to extend the
formalism to four-dimensional Einstein-Cartan gravity, where the symplectic form is degenerate. We proceed with a
generic diffeomorphism invariant field theory, with general relativity as the goal in mind.

The base manifold on which dynamics occurs is an $n$-dimensional differentiable manifold $M^{*}$, which represents the 
entire dynamical arena. In $M^{*}$ we embed an $(n-1)$-dimensional hypersurface $\Sigma$ by the embedding map
$\sigma:\Sigma\rightarrow M^{*}$. 
Although it is likely not necessary in the most general context, to
make contact with the ordinary Hamiltonian framework, we will assume that $\Sigma$ is a spacelike hypersurface, which
is a Cauchy slice when $M^{*}$ is Reimannian, and the global topology is fixed to be homotopic to $\mathbb{R}\times\Sigma$.
The arena that we are concerned with is the portion of spacetime including all points in $\Sigma$ and all points in the
past of
$\Sigma$. Thus, we embed a new manifold $M$ in $M^{*}$ such that $\partial M=\sigma(\Sigma)$ and $M-\partial M$ contains
all
and only points in the past of $\Sigma$. Thus, in
this picture, the manifold $M$ evolves by pushing $
\partial M$ forward along a future directed timelike vector field. We will occasionally be loose and simply write $\partial
M=\Sigma$. The embedding of $\Sigma$ in $M$ also induces a projection map, $h:\Gamma_{L}\rightarrow \Gamma_{H}$, of the
Lagrangian phase space down to the Hamiltonian phase space. If $f$ is a functional on $\Gamma_{H}$, its pullback under
this projection, $h^{*}f$, is a functional that depends only on the boundary data and is independent of variations in
the bulk. For simple integral functionals, which we will primarily be concerned with in this paper, there is an obvious projection from functionals on $\Gamma_L$ to $\Gamma_H$. Given a functional on $\Gamma_L$ that can be written in the form $f=\int_M d\widetilde{\alpha}$, for some $(n-1)$--form $\widetilde{\alpha}$, then $h(f)=\int_{\Sigma}\sigma^*\widetilde{\alpha}$, where $\sigma^*\widetilde{\alpha}$ is the pull-back of $\widetilde{\alpha}$ to $\Sigma$. 

The starting point for our formalism is the action on $M$:
\beq
S=\int_{M}\widetilde{L}
\eeq
where the Lagrangian 4-form, $\widetilde{L}$, is a functional of a finite set of dynamical fields, $\Phi^{(i)}$. The
dynamical fields form a set of coordinates on the Lagrangian phase space, $\Gamma_{L}$. The points of the Hamiltonian
phase space, $\Gamma_{H}$, are configurations of the boundary data which can be coordinatized by the pull-back fields
$\phi^{(i)}\equiv (\sigma^{*}\Phi^{(i)},\sigma^{*}d\Phi^{(i)})$.
An arbitrary
variation of the action is given by $\bm{\iota_{\bm{\bar{X}}}}\dl S$, where $\dl$ is the exterior derivative on the function
space, $\Gamma_{L}$, and $\bm{\bar{X}}$ is a vector field in $T\Gamma_{L}$. In general, an arbitrary variation will
have a boundary piece and a bulk piece:
\beqa
\dl S &=& (\dl S)_{boundary}+(\dl S)_{bulk} \nn\\
&=& \bm{J}+\bm{\theta}.
\eeqa
As previously, we have defined the symplectic one form $\bm{J}\equiv (\dl S)_{boundary}$, and
the Lagrangian one form, $\bm{\theta}\equiv (\dl S)_{bulk}$. The symplectic one form, or symplectic
potential, plays an important role in many generalizations of classical and quantum Hamiltonian mechanics, most
notably geometric quantization where it plays the role of a $U(1)$ connection on a complex line bundle (the
pre-quantum Hilbert space) over the Hamiltonian phase space. We will use this interpretation in the last section when 
we discuss the quantum theory. The Lagrangian one-form is particularly useful in that the solutions to the bulk
equations of motion lie in
the submanifold of $\Gamma_{L}$ where $\bm{\theta}\approx 0$. For now, the most salient feature of the symplectic
and Lagrangian one-forms is that their exterior derivatives, as in the previous section, give the standard symplectic
form on the classical Hamiltonian phase space:
\beq
\bm{\Omega}\equiv -\dl \bm{J}=\dl\bm{\theta}\label{SymIdentity}
\eeq
where we have used the identity $\dl\dl S=\dl\bm{J}+\dl\bm{\theta}=0 $.
As such, the (pre)symplectic form may be degenerate simply by definition, even when it is not restricted to any solution submanifold\footnote{In this paper we will be loose and often use the term symplectic (as opposed to pre-symplectic) to refer to a closed but possibly degenerate two-form}.

We emphasize here, whereas $\bm{J}$ depends only on the boundary variables and can identified with the pull-back of a
one--form on $\Gamma_{H}$,
$\bm{\theta}$ is an $n$-dimensional object, whose variation happens to be a pure boundary term, namely the symplectic
form. Given this identity, we will have little use for the symplectic one-form, formulating Hamiltonian
dynamics entirely in terms of the Lagrangian one-form, $\bm{\theta}$.

\subsection{Hamiltonian functionals and Noether vectors}
Suppose the vector field $\bm{\bar{X}}$ was the generator of a symplectomorphism so that $\bm{\mathcal{L}_{\bar{X}}\Omega}=0$. This implies that at least locally (globally if the cohomology class is trivial) one can find a functional $X$ such that $\bm{\Omega(\bar{X},\ )}=\dl X$. Now, consider the Lie derivative of the Lagrangian one--form along any vector field $\bm{\bar{X}}$ given by
$\bm{\mathcal{L}_{\bar{X}} \theta}=\bm{\iota_{\bar{X}}\dl \theta}+\dl \bm{\iota_{\bar{X}}\theta}$. From the
identity (\ref{SymIdentity}), rearranging terms we have
\beq
\bm{\Omega(\bar{X},\ )}=\dl X=\bm{\mathcal{L}_{\bar{X}}\theta}-\dl \bm{\theta(\bar{X})} .\label{SymIdentity2}
\eeq
It is clear that $\bm{\mathcal{L}_{\bar{X}}\theta}$ is (locally) an exact form, so we can write
$\bm{\mathcal{L}_{\bar{X}}\theta}=\dl f$ with $X=f-\bm{\theta(\bar{X})}$.
The association is unique only up to addition of a constant to $g$ and addition of a vector in the kernel of $\bm{\Omega}$
to $\bm{\bar{X}}$.

The special set of of functions of the form $X=-\bm{\theta(\bar{X})}$ will play an important role in what
follows. Such functions have the peculiar property that they must vanish
wherever $\bm{\theta}$ vanishes, so not every functional is of this form. Given any arbitrary functional $\alpha$, consider the new functional
$Y=\alpha
(-\bm{\theta(\bar{X})})=-\bm{\theta(}\alpha \bm{\bar{X} )}$. From the second equality it is clear that $Y$ is
also of the form $Y=-\bm{\theta(\bar{Y})}$. But,
since $\alpha$ is arbitrary, the only restriction on $Y$ is that it must vanish
(smoothly if we demand continuity and differentiability) on the submanifold where $\bm{\theta}\approx 0$. But this is
precisely the submanifold where the bulk equations of motion hold. Thus, these functionals will play an important role in defining the classical phase space.

We now consider a special class of such functions that are related to the symmetries of the classical action.
Consider a vector field $\bm{\bar{W}}$ that satisfies
\beq
-\bm{\theta(\bar{W})}\equiv C_{W}={\textstyle boundary\  functional}.
\eeq
We will refer to such vector fields as Noether vectors (not to be confused with Noether currents) since they generate a
change in the action that vanishes in the bulk. The remaining boundary functional must vanish on solutions of the equations of motion. But, since the functional only depends on boundary data, the restriction of the functional to the submanifold where the equations of motion hold can been seen as a constraint on the initial data. Thus, Noether vectors are naturally associated with boundary constraints. As an example, we
consider a one parameter diffeomorphism generated by $\bar{V}$
in $TM^{*}$. The associated field $\bm{\bar{V}}=\int_{M}\mathcal{L}_{\bar{V}}\Phi^{(i)}\,\frac{\dl}{\dl\Phi^{(i)}}+\mathcal{L}_{\bar{V}}d\Phi^{(j)}\,\frac{\dl}{\dl d\Phi^{(j)}}$ is a Noether
vector since we have assumed that the action is diffeomorphism invariant in the bulk. The resulting constraint
$C(\bar{V})\equiv -\bm{\theta(\bar{V})}$ is a boundary functional that generates diffeomorphisms on the boundary.
The restriction to the constraint submanifold $C(\bar{V})\approx 0$ is a restriction on the admissible initial data. Later on, we will construct such constraints explicitly for the case of Einstein-Cartan gravity.

\subsection{Hamilton's equations}
We are now in a position to discuss Hamilton's equations in this formalism. As previously the Lagrangian is a functional of the dynamical fields $\Phi^{(i)}$, which we will assume are differential forms, and their derivatives $d\Phi^{(j)}$. To this end, we define a monotonically
increasing time
function, $t$, on $M^{*}$ such that the boundary Cauchy slice, $\Sigma$, occurs at $t=t_0$. On $\Sigma$, we define a set of
one-forms,
$dt$ and $dx^{a}$, where $x^{a}$ are spatial coordinates on $\Sigma$. Next we choose a timelike vector $\bar{t}\in
TM^{*}$,
such that $dt(\bar{t})=1$ and $dx^{a}(\bar{t})=0$. Given such a choice for the time evolution vector field, consider
its associated Noether vector
$\bm{\bar{t}}=\int_{M}\mathcal{L}_{\bar{t}}\Phi^{(i)}\frac{\dl}{\dl\Phi^{(i)}}+\mathcal{L}_{\bar{t}}d\Phi^{(j)}\frac{\dl}{\dl d\Phi^{(j)}}$. This vector is Noether since the action 
is invariant in the bulk under four-dimensional diffeomorphisms. Following example from the case of the non-relativistic
particle discussed previously, we guess that Hamilton's equations can be summarized in the form
\beq
\bm{\mathcal{L}_{\bm{\bar{t}}}\,\theta} = 0 \,.\label{Ham0}
\eeq
As we will see, this equation is a restriction only on the boundary data. The Hamiltonian phase space consists of functionals that depend only on the pull-back fields and possibly their derivatives $\phi^{(i)}\equiv(\sigma^*\Phi^{(i)},\sigma^* d\Phi^{(i)})$. The subspace of $\Gamma_{H}$ where
(\ref{Ham0}) holds we will refer to as the constraint submanifold $\bar{\Gamma}_{H}$.

To see that this reduces to the ordinary form of Hamilton's equations we first note that from the the identity
(\ref{SymIdentity}), equation (\ref{Ham0}) yields:
\beq
\bm{\Omega(\bar{t},\ )}=\dl(-\bm{\theta(\bar{t})})\,. \label{Ham0a}
\eeq
Since $\bm{\bar{t}}$ generates a diffeomorphism along $\bar{t}$, and the theory is assumed to be diffeomorphism invariant, $-\bm{\theta(\bar{t})}$ is a pure boundary term. Specifically, it is given by
$-\bm{\theta(\bar{t})}=\bm{J(\bar{t})}-\int_{\Sigma}\widetilde{L}(\bar{t})$. This boundary functional may depend on field components denoted $\lambda^{(i)}$ that, despite being ordinary dynamical variables on $\Gamma_H$, do not occur in the symplectic form or in the left hand side of (\ref{Ham0a}). In general the gradient $\dl\bm{(\theta(\bar{t}))}$ on $\Gamma_H$ 
will contain terms of the form $\bm{\theta(\bar{t})}[\dl
\lambda^{(i)}]$.
On the other hand, since the left hand side of (\ref{Ham0a})
is independent of the $\lambda^{(i)}$, we must have
$\bm{\theta(\bar{t})}[\dl\lambda^{(i)}]= 0$. Since the variations
$\dl\lambda^{(i)}$ are arbitrary, in particular we have $\bm{\theta(\bar{t})}= 0$ for all configurations $\lambda^{(i)}$. The remaining terms
can be viewed as
functionals on $\Gamma_{H}$ that depend on an arbitrary choice of smearing functions
$\lambda^{(i)}$. We
can then summarize
(\ref{Ham0}) as follows:
\beqa
\bm{HAM1:} &\quad & \bm{\theta(\bar{t})}= 0 \ \ \ \ \  \forall \lambda^{(i)} \label{Ham1}\\ 
\bm{HAM2:} &\quad & \bm{\Omega(\bar{t},\ )}=\dl(-\bm{\theta(\bar{t})})\Big|_{\lambda^{(i)}} \label{Ham2}
\eeqa
where the notation on the right hand side of the second line indicates that the gradient in this expression is to be
taken holding
$\lambda^{(i)}$
fixed. 
This formulation of Hamilton's equations are likely unfamiliar so some discussion is in order. The first equation,
HAM1,
is simply the statement that we are restricting ourselves to a submanifold where diffeomorphisms along $\bar{t}$ are an
exact symmetries in the bulk and on the boundary. As we will see explicitly for general relativity,
 $-\bm{\theta(\bar{t})}=C_{tot}(\lambda^{(i)})$ is the
total Hamiltonian
constraint so this simply defines the constraint submanifold $\Gamma^{(1)}_{H}\subset \Gamma_{H}$. The second
equation, HAM2, then takes the form
\beq
\bm{\Omega(\bar{t},\ )}=\dl C_{tot}(\lambda^{(i)})\Big|_{\lambda^{(i)}}
\eeq
which is the ordinary form of Hamilton's equations given the total Hamiltonian constraint.
This equation has two non-trivial implications. First, it shows that there must exist a canonical vector field
associated with the functional $-\bm{\theta(\bar{t})}$. We
recall that for a non-degenerate symplectic form this is not always true.
Thus in contrast to the ordinary Hamiltonian formalism based on a non-degenerate symplectic form, HAM2 itself
defines a constraint submanifold, which we denote by $\Gamma^{(2)}_{H}\subset \Gamma_{H}$. The true submanifold on which
the Hamiltonian dynamics occurs is then $\bar{\Gamma}_{H}=\Gamma^{(1)}_{H}\bigcap \Gamma^{(2)}_{H}\subset \Gamma_{H}$.
Second, HAM2 shows that the total Hamiltonian constraint does indeed generate the time evolution of the system. 
 
\section{An application: general relativity} 
Hamiltonian general relativity can naturally be cast in the formalism developed above. Since the theory is
diffeomorphism invariant, the time evolution of the dynamical variables is generated by a Noether vector. In addition,
by allowing for torsion, the action can be written as a first order theory. This means that the Hamiltonian variables
can be viewed as simply the pull-backs of the Lagrangian variables (with no derivatives) to the spacelike Cauchy surface, thereby avoiding
the necessity of performing a Legendre transformation in order to make contact with the ordinary Hamiltonian
formalism. 

The dynamical variables in the Einstein-Cartan formalism are the $spin(3,1)$-valued connection coefficients, $\varpi$,
and a frame field or
tetrad, $\vep$, which is a one-form taking values in the adjoint representation of $Spin(3,1)$.\footnote{For
definiteness, we can use a Clifford notation where $\varpi=\frac{1}{4}\gamma_{[I}\gamma_{J]}\,\varpi^{IJ}$ and
$\vep=\frac{i}{2}\gamma_{I}\,\vep^{I}$. In this notation, the dual operator is given by
$\star=-i\gamma^{5}=\gamma^{0}\gamma^{1}\gamma^{2}\gamma^{3}$. For simplicity, in the integral we will drop the
explicit trace over the Dirac matrices, and the the wedge product between forms will be assumed. When the wedge
product is written explicitly it will generally be the wedge product on the phase space.}
It will be convenient to distinguish the four-dimensional variables on $M$ from the three-dimensional 
variables on $\Sigma$. Thus, we define the pull-back of the dynamical fields, $e\equiv\sigma^{*}\vep$ and
$\omega\equiv\sigma^{*}\varpi$.
The curvature of the connection is $\mathcal{R}=d\varpi+\varpi\,\varpi$ and the curvature of the pull-back connection is
$R=d\omega+\omega\,\omega$. We
begin with the Einstein-Cartan action:
\beq
S=\frac{1}{k}\int_{M}\star \,\vep\,\vep\,\mathcal{R}\,.
\eeq
Taking the gradient we have
\beqa
\dl S&=&\frac{1}{k}\int_{M}\star \,\vep\,\vep\,D_\varpi\dl\varpi +(\star \mathcal{R}\,\vep-\vep\star\! \mathcal{R})\dl \vep\nn\\
&=&\frac{1}{k}\int_{M}d\left(\star \,\vep\,\vep\,\dl \varpi\right) +\frac{1}{k}\int_{M}-D_\varpi(\star \,\vep\,\vep)\dl\varpi+(\star
\mathcal{R}\,\vep-\vep\star
\!\mathcal{R})\,\dl \vep\,.\nn
\eeqa
From this we immediately identify the symplectic and Lagrangian one-forms:
\beqa
\bm{J}&=&\frac{1}{k}\int_{M}d\left(\star \,\vep\,\vep\,\dl \varpi\right)
=\frac{1}{k}\int_{\partial M=\Sigma}\star \,e\,e\,\dl
\omega \\
\bm{\theta}&=&\frac{1}{k}\int_{M}-D_\varpi (\star\vep\,\vep)\dl\varpi+(\star \mathcal{R}\,\vep-\vep\star
\!\mathcal{R})\,\dl \vep\,.
\eeqa
The symplectic form is then
\beq
\bm{\Omega}=-\dl\bm{J}=\dl\bm{\theta}=\frac{1}{k}\int_{\Sigma}\star\dl\omega\wedge\dl(e\,e)\,.
\eeq
This symplectic form is degenerate
since the ``momentum", $\star\, e\,e\,$, is not an arbitrary (bi-vector valued) two-form. We stress that the degeneracy does not emerge from the restriction of the symplectic form to a constraint submanifold as is usually the case. Nor is the degeneracy of the same character as that of the pre-symplectic form on the covariant phase space discussed in section (\ref{StandardCovariant}). In the present case, the degeneracy is a peculiarity of Einstein-Cartan gravity---it is not a generic feature of the method we have used to construct the symplectic form.

The bulk equations of motion $\bm{\theta}=0$ are the ordinary Einstein-Cartan equations of motion:
\beqa
D_\varpi(\star \,\vep\,\vep)&=& 0\label{ECEOM1} \\ 
\star \mathcal{R}\,\vep-\vep\star\! \mathcal{R} &=& 0 \label{ECEOM2}
\eeqa

There are two important Noether vectors that are immediately relevant to the Hamilton field equations. First we note
that the action is invariant in the bulk under local $Spin(3,1)$ gauge transformations. The Noether vector corresponding
to an infinitesimal gauge transformation generated by $\lambda\in so(3,1)$ is 
\beq
\bm{\bar{\lambda}}=\int_{M}-D_\varpi\lambda\,\frac{\dl}{\dl\varpi}+[\lambda,\vep]\,\frac{\dl}{\dl \vep}\,.
\eeq
The corresponding constraint functional, which for obvious reasons we will refer to as the Gauss constraint, is 
\beqa
C_{G}(\lambda)&\equiv& -\bm{\theta(\bar{\lambda})}\nn\\
&=& \frac{1}{k}\int_{M}D_\varpi\lambda\,D_\varpi(\star \,\vep\,\vep)+[\lambda,\vep](\star \mathcal{R}\,\vep-\vep\star \!\mathcal{R})\nn\\
&=& \frac{1}{k}\int_{\partial M} -D_\omega\lambda\star e\,e\,.
\eeqa
In addition, the bulk action is invariant under diffeomorphisms. Thus, let $\bar{V}$ be any four-vector on $M^{*}$. It
generates an infinitesimal one-parameter diffeomorphism along its integral lines. The corresponding Noether vector is
\beq
\bm{\bar{V}}=\int_{M}\mathcal{L}_{\bar{V}}\varpi\,\frac{\dl}{\dl\varpi}+\mathcal{L}_{\bar{V}}\vep\frac{\dl}{\dl \vep}
\eeq 
and the corresponding constraint is 
\beqa
-\bm{\theta(\bar{V})}&=&\frac{1}{k}\int_{M}-\mathcal{L}_{\bar{V}}\varpi\,D_\varpi(\star\, \vep\,\vep)
+\mathcal{L}_{\bar{V}}\vep(\star \mathcal{R}\,\vep-\vep\star \!\mathcal{R})\nn\\
&=&\frac{1}{k}\int_{\partial M}\mathcal{L}_{\bar{V}}\varpi\star \vep\,\vep-\iota_{\bar{V}}(\star \vep\,\vep\,\mathcal{R})\nn\\
&=&\frac{1}{k}\int_{\partial M}-\star [\vep(\bar{V}),e]\,R+D_\omega(\varpi(\bar{V}))\star e\,e\,.\label{HamConstraint1}
\eeqa
Now, given a choice of coordinates $(t, x^{a})$ and there associated one-forms $(dt, dx^{a})$,
we can divide $\bar{V}$ into components perpendicular and parallel to
$\Sigma$. First suppose $\bar{V}$ is everywhere parallel to the boundary manifold $\partial M$ and call this boundary vector $\bar{N}$.
In this case $\bar{N}$ generates a three dimensional diffeomorphism on the boundary. Then the constraint reduces to
the ordinary diffeomorphism constraint:
\beq
C_{D}(\bar{N})=\int_{\partial M}\star \mathcal{L}_{\bar{N}}\omega \,e\,e\,.
\eeq
Now consider the case where $\bar{V}$ is normal to the boundary and timelike. Denote this vector by $\bar{t}$. Then
the constraint functional is what we have referred to as the total Hamiltonian $C_{tot}$, which we can now construct
explicitly:
\beqa
C_{tot}&=&C_{H}(t)+C_{G}(\lambda)\nn\\
&=&\frac{1}{k}\int_{\partial M}-\star[t,e]\,R-D_\omega \lambda \star e\,e \label{HamConstraint2}
\eeqa
where we have defined $t\equiv \vep(\bar{t})$ and $\lambda\equiv-\varpi(\bar{t})$, and noticed that part of the
Hamiltonian
is identical to the Gauss constraint. The remaining piece $C_{H}(t)=-\frac{1}{k}\int_{\partial M}\star[t,e]\,R$ 
we will refer to as the Hamiltonian constraint. As
opposed to most standard Hamiltonian formulations, the Hamiltonian constraint here is not a scalar constraint; rather,
it is vectorial since $t=\frac{i}{2}\gamma_{I}t^{I}$ has four independent components that transform like a four-vector
under $Spin(3,1)$. Loosely speaking, the vectorial Hamiltonian contains both the ordinay scalar and the diffeomorphism
constraints, however, it is such that one cannot split the vectorial Hamiltonian into these two components without
breaking the gauge. We will have more to say on the nature of this constraint and its close relation to the de Sitter
group in a follow-up paper \cite{Randono:CovariantCanonical}. 

\subsection{Hamilton's equations for Einstein-Cartan gravity}
Let us now consider Hamilton's equations (\ref{Ham0}) for this theory. From the identity (\ref{SymIdentity}), this can
be rewritten
\beq
\bm{\Omega(\bar{t},\ )}=\dl(-\bm{\theta(\bar{t})})\label{Ham0b}
\eeq
where it is recognized that the variation includes the the terms $\dl{\lambda}=-\iota_{\bar{t}}\dl\varpi$ and $\dl t=\iota_{\bar{t}}\dl \vep$.
From the explicit expression for the symplectic form, $\bm{\Omega}=\frac{1}{k}\int_{\Sigma}\star \dl \omega \wedge \dl (e\,e)$,
the left hand side is
\beq
\bm{\Omega(\bar{t}, \ )}=\int_{\Sigma}(e \star\! \mathcal{L}_{\bar{t}}\omega 
+\star \mathcal{L}_{\bar{t}}\omega \, e)\, 
\dl e - \star\left(\mathcal{L}_{\bar{t}}e\, e +e\, \mathcal{L}_{\bar{t}}e\right)\, \dl \omega\,.
\eeq
To compute the right hand side we need to compute the exterior derivative of the constraints:
\beqa
\dl C_{G}(\lambda)&=&\frac{1}{k}\int_{\Sigma}-[\lambda, \star\,e\,e]\,\dl\omega
-\left(e \star \! D_\omega\lambda+\star D_\omega\lambda\,e\right)\, \dl e -D_\omega\dl\lambda \star e\,e\nn\\
\dl C_{H}(t)&=& \frac{1}{k}\int_{\Sigma}-\,D_\omega[t, e]\,\dl\omega  +[t, \star R]\,\dl e-\star[\dl t,e]\,R\,.
\eeqa
Since the left hand side is independent of $\dl{\lambda}$
and $\dl t$, we must have
\beqa
C_{tot}(\dl t,\dl\lambda) &=& C_{H}(\dl t)+C_{G}(\dl \lambda)\nn\\
&=& \frac{1}{k}\int_{\Sigma}-\star[\dl t,e]\,R-D\dl\lambda \star e\,e \approx 0\,.
\eeqa
The variations are arbitrary so the above reduces to the pointwise equations:
\beqa
D_{\omega}(\star \,e\,e)=\sigma^{*}\left(D_{\varpi}(\star \,\vep\,\vep)\right)&\approx &0 \label{EE1a}\nn\\
\star R\,e-e\star \!R=\sigma^{*}\left(\star \mathcal{R}\,\vep-\vep\star \!\mathcal{R}\right)
 &\approx & 0\,.\label{EE1b}
\eeqa
These equations are clearly the pull-backs of the Einstein-Cartan equations of motion (\ref{ECEOM1}) and (\ref{ECEOM2})
to the boundary. Thus, it is clear that HAM1 for general relativity is simply the spatial components of the Lagrangian
field equations on the boundary. We note that three dimensional diffeomorphism invariance is already
implicitly contained in the above equations. To see this, we recall the identity
$C_{D}=C_{H}(e(\bar{N}))+C_{G}(-\omega(\bar{N}))$ from which it is clear that $C_{D}(\bar{N})\approx 0$ on
$\Gamma^{(1)}_{H}$.

To compute HAM2 , we identify the left and right hand sides of the remaining components of (\ref{Ham0b}) for arbitrary
variations, $\dl\omega$ and $\dl e$:
\beqa
\star(\mathcal{L}_{\bar{t}}e\, e +e\, \mathcal{L}_{\bar{t}}e) &=& [\lambda,
\star\,e\,e] +\star\,D_\omega[t, e] \label{EE2a}\\
e \star\! \mathcal{L}_{\bar{t}}\,\omega +\star \mathcal{L}_{\bar{t}}\,\omega \, e
&=&-\left(e \star D_\omega\lambda+\star\,D_\omega\lambda\,e\right)+[t, \star\,R] \label{EE2b}
\eeqa
These complicated looking expressions can, in fact, be deciphered rather easily. The first equation, (\ref{EE2a}), is
precisely
pull-back of the time component of the bulk equation of motion,
(\ref{ECEOM1}),
\beq
\sigma^{*}\left(\iota_{\bar{t}}\left(D_{\varpi}(\star \,\vep\,\vep)\right)\right)=0
\eeq
and the second set, (\ref{EE2b}), is the time component of the bulk equation of motion, (\ref{ECEOM2}),
\beq
\sigma^{*}\left(\iota_{\bar{t}}\left(\star \mathcal{R}\, \vep -\vep\star\! \mathcal{R}\right) \right)= 0 \,.
\eeq

Thus, we have seen explicitly that Hamilton's equations (\ref{Ham0}) are precisely the
Einstein-Cartan equations of motion pulled back to the boundary $\Sigma=\partial M$. All
solutions to the boundary equations of motion are points of $\bar{\Gamma}_{H}=\Gamma^{(1)}_{H}\bigcap
\Gamma^{(2)}_{H}$, and all points of $\bar{\Gamma}_{H}$ are solutions to the boundary equations of motion.

\section{The constraint algebra}
Given a vector field that preserves the symplectic form in the sense that
$\bm{\mathcal{L}_{\bar{X}}}\bm{\Omega}=\dl(\bm{\Omega(\bar{X},\ )})=0$,
we can always locally construct a function, $X$, such that $\bm{\Omega(\bar{X},\ )}=\dl X$.
However, since the symplectic form is degenerate the pairing is not unique. Given another vector field
$\bm{\bar{X'}}$ that also satisfies $\bm{\Omega(\bar{X'},\ )}=\dl X$, we have $\bm{\Omega(\bar{X'}-\bar{X},\
)}=0$ so that $\bm{\Delta\bar{X}}=\bm{\bar{X'}}-\bm{\bar{X}}$ must be in the kernel of $\bm{\Omega}$. Given two such
functions, we can define the Poisson bracket by
\beq
\{X,Y\}\equiv \bm{\Omega(\bm{\bar{X}}, \bm{\bar{Y}})}\,.
\eeq
Despite the non-uniqueness of the canonical vector fields, the Poisson bracket is unique since 
\beqa
\bm{\Omega(\bm{\bar{X'}},\bm{\bar{Y'}})}
&=&\bm{\Omega(\bm{\bar{X}}+\Delta\bm{\bar{X}}, \bm{\bar{Y}}+\Delta\bm{\bar{Y}})}\nn\\
&=&\bm{\Omega}(\bm{\bar{X}},\bm{\bar{Y}})
\eeqa
where we have used the fact that $\bm{\Delta\bar{X}}$ and $\bm{\Delta\bar{Y}}$ are in the kernel of $\bm{\Omega}$.
Furthermore, one can still associate a Hamiltonian vector field with the functional $\{X,Y\}$ and the vector field is
(up to addition of a vector in the kernel) the Lie bracket $\bm{[\bm{\bar{Y}},\bm{\bar{X}}]}$:
\beq
\bm{\Omega([\bm{\bar{Y}},\bm{\bar{X}}],\ )}=\dl\{X,Y\}\,. 
\eeq
Thus, despite the non-degeneracy of the symplectic form, given two generating functionals $X$ and $Y$, the Poisson
bracket can be defined and it satisfies all the usual properties of a Poisson bracket. The only difference is that not
all functionals are generating functionals.

Now consider the expression $\{X_{\bm{\bar{X}}},Y_{\bm{\bar{Y}}}\}_{\bm{\theta}}\equiv
\bm{\theta([\bm{\bar{X}},\bm{\bar{Y}}])}$.\footnote{It may be tempting to think of 
$\{X_{\bm{\bar{X}}},Y_{\bm{\bar{Y}}}\}_{\bm{\theta}}$ as a natural generalization of the Poisson bracket. The bracket is
obviously anti-symmetric and one can easily show that it satisfies the Jacobi identity since the Lie bracket between
vector fields satisfies the Jacobi identity. However, the bracket is not Leibnitz. This problem may be overcome
by considering a modified scalar product, however, we will not consider such a generalization here. Furthermore, the
bracket is not unique to the two functions involved, but also depends on the choice of associated vector fields.
Again, this problem could potentially be overcome by considering the mapping between the cohomology class of the
function $X$ and the degeneracy class of a the associated vector field $\bm{\bar{X}}$.}
Using the identity
\beq
\bm{\theta([\bm{\bar{X}},\bm{\bar{Y}}])}=\bm{\Omega(\bm{\bar{X}},\bm{\bar{Y}})}
+\bm{\iota}_{\bm{\bar{X}}}\bm{\mathcal{L}}_{\bm{\bar{Y}}}\bm{\theta}-\bm{\iota}_{\bm{\bar{Y}}}\bm{\mathcal{L}}_{\bm{\bar{X}}}\bm{\theta}
\eeq
we see that $\{X_{\bm{\bar{X}}},Y_{\bm{\bar{Y}}}\}_{\bm{\theta}}\approx \{X,Y\}_{Poisson}$ precisely when the analogue
of Hamilton's equation (\ref{Ham0}), $\bm{\mathcal{L}}_{\bm{\bar{X}}}\bm{\theta}=\bm{\mathcal{L}}_{\bm{\bar{Y}}}\bm{\theta}=0$,
holds. This will allow us to compute the commutator of the constraints on $\bar{\Gamma}_{H}$. 

\subsection{The constraint algebra of general relativity}
Some commutators are easier to compute by working with $\bm{\theta([\bm{\bar{X}},\bm{\bar{Y}}])}$ directly rather than
$\bm{\Omega(\bm{\bar{X}},\bm{\bar{Y}})}$. We give an example here. Returning to Einstein-Cartan gravity, suppose we
have two different choices for the time evolution vector field, $\bar{t}_{1}$ and $\bar{t}_{2}$. Both generate
diffeomorphisms along a timelike vector field and their corresponding Noether vectors are
\beqa
\bm{\bar{t}_{1,2}}&=&\int_{M}\mathcal{L}_{\bar{t}_{1,2}}\vep\,\frac{\dl}{\dl
\vep}+\mathcal{L}_{\bar{t}_{1,2}}\varpi\,\frac{\dl}{\dl
\varpi}
\eeqa
The commutator between the two Noether vectors is
\beq
\bm{[\bm{\bar{t}_{1}},\bm{\bar{t}_{2}}]}=\int_{M}\mathcal{L}_{[\bar{t}_{1},\bar{t}_{2}]}\vep\,\frac{\dl}{\dl \vep}
+\mathcal{L}_{[\bar{t}_{1},\bar{t}_{2}]}\varpi\,\frac{\dl}{\dl\varpi}\,.
\eeq
Writing $\bar{T}\equiv -[\bar{t}_{1},\bar{t}_{2}]$, we see from (\ref{HamConstraint1}) and (\ref{HamConstraint2}) that
\beq
\bm{\theta([\bm{\bar{t}_{1}},\bm{\bar{t}_{2}}])}=C_{H}(\vep(\bar{T}))+C_{G}(-\varpi(\bar{T}))\,.
\eeq
Since both $\bar{t}_{1}$ and $\bar{t}_{2}$ are good time variables, Hamilton's equations, apply to both. Thus on
$\bar{\Gamma}_{H}$ we have $\bm{\mathcal{L}}_{\bm{\bar{t}_{1}}}\bm{\theta}=\bm{\mathcal{L}}_{\bm{\bar{t}_{2}}}\bm{\theta}=0$.
In this case we have
\beqa
\bm{\theta([\bm{\bar{t}_{1}},\bm{\bar{t}_{2}}])}&\stackrel{\bar{\Gamma}_{H}}{\approx}&
\{C_{tot}(t_{1},\lambda_{1}),C_{tot}(t_{2},\lambda_{2})\}\nn\\
&=& \{C_{H}(t_{1})+C_{G}(\lambda_{1})\,,\,C_{H}(t_{2})+C_{G}(\lambda_{2})\}\nn\\
&=& C_{H}(\vep(\bar{T}))+C_{G}(-\varpi(\bar{T}))\,.
\eeqa
If one chooses a normal vector to the foliation, one can then divide vectors into components perpendicular and parallel to the foliation so that  $\bar{T}=\bar{T}_{\|}+\bar{T}_{\perp}$. The above result then reduces to the following 
\beqa
\{C_{tot}(t_{1},\lambda_{1})\,,\,C_{tot}(t_{2},\lambda_{2})\}=C_D(\bar{T}_{\|})+C_H(T)+C_G(\Lambda)
\eeqa
where $T\equiv\vep(\bar{T}_{\perp})$ and $\Lambda\equiv-\varpi(\bar{T}_{\perp})$. Since all of these constraints vanish on the constraint
submanifold $\Gamma^{(1)}_{H}$, we immediately see
\beq
\{C_{tot}(t_{1},\lambda_{1}),C_{tot}(t_{2},\lambda_{2})\}=\{C_{H}(t_{1})+C_{G}(\lambda_{1})\,,\,C_{H}(t_{2})+C_{G}(\lambda_{2})\}
\stackrel{\bar{\Gamma}_{H}}{\approx} 0\,.
\eeq
This implies that the time evolution of dynamical fields on $\bar{\Gamma}_{H}$ does not pull us off the submanifold,
which in turn, means the constraints and the Hamiltonian evolution equations are self-consistent. We emphasize that we
have not had need to gauge fix the action or mod out by the orbits of the kernel of $\bm{\Omega}$.

In a follow-up paper we compute the constraint algebra explicitly for the case of a positive cosmological constant and
show that the algebra is a deformation of the de Sitter, anti-de Sitter, or Poincar\'{e} algebras depending on the
value of the cosmological constant, and the deformation is the Weyl tensor \cite{Randono:CovariantCanonical}. This
illuminates the vectorial nature of the Hamiltonian constraint---in an appropriate limit, the vector generators of the
Hamiltonian constraint must be the translation generators of the $(A)dS$ or Poinar\'{e} group.
In addition we will discuss possible routes to quantization and the relevance of this formalism to the Kodama state.

\section{Quantization}
We now consider a possible route to quantization in the framework developed above. We will begin with a very brief review of the pre-quantization procedure in the standard representation \cite{Nair:QFT, Woodhouse:GQ, Sniatycki:GQ}. 

The key insight of geometric quantization is the identification of the symplectic one-form, $\bm{J}$, as a $U(1)$ connection on a complex line bundle over the Hamiltonian phase space which comprises the ``pre-quantum" Hilbert space. We first review the procedure of constructing states and operators on the pre-quantum Hilbert space. Since we are dealing with a degenerate (pre)symplectic form, not all functionals can be lifted to quantum operators. For this reason, we will focus on generators of pre-symplectomorphisms, which can always locally be associated with a boundary functionals that can be lifted to pre-quantum operators. Thus, consider a (pre)symplecto-morphism generated by a vector field $\bm{\bar{X}}$, so that $\bm{\mathcal{L}_{\bar{X}}\Omega}=0$. Since $\bm{\Omega}$ is closed ($\bm{\mathcal{L}_{\bar{X}}\Omega}=\dl(\bm{(\Omega(\bar{X},\ )})=0$), so $\bm{\Omega(\bar{X},\ )}$ is exact. This means that at least locally (globally if the cohomology class is trivial) we can pair the vector field $\bm{\bar{X}}$ with a function $X$, by the identification $\bm{\Omega(\bar{X},\ )}=\dl X$. Now, consider the change in the symplectic one-form under the same symplecto-morphism:
\beqa
\bm{J} & \rightarrow & \bm{J}+\bm{\mathcal{L}_{\bar{X}}J}\nn\\
&=&\bm{J}+\dl(\bm{J(\bar{X})}-X)\nn\\
&=&\bm{J}+\dl\Lambda\,.
\eeqa
Where in the last line we have defined $\Lambda \equiv \bm{J(\bar{X})}-X$. We now identify $\bm{J}$ with a $U(1)$ connection coefficient associated with the covariant exterior derivative $\bm{\mathcal{D}_J}\equiv \dl-i\bm{J}$. With this identification, we recognize the above transformation as a generic change in a connection under the gauge transformation generated by $g=e^{i\Lambda}$, given by $-i\bm{J}\rightarrow g(-i\bm{J})g^{-1}-(dg)g^{-1}$. Thus, the symplecto-morphism plays a dual role on the $U(1)$ bundle. First, it is a diffeomorphism, and second, it can also be identified with a $U(1)$-gauge transformation.

Next we augment the phase space with a complex line bundle on which the connection acts, and consider an arbitrary section of the bundle $\psi$. We refer to $\psi$ as the ``pre-quantum" wave function. As such it depends on all the phase space variables, so in the case of Einstein-Cartan gravity, $\psi=\psi[\omega, e]$. The infinitesimal transformation of the pre-quantum wavefunction is a combination of both a diffeomorphism and a $U(1)$ gauge transformation:
\beqa
\psi&\rightarrow& \psi+\bm{\mathcal{L}_{\bar{X}}}\psi+\beta\,i\Lambda\,\psi \nn\\
&=& \psi+\left(\bm{\iota_{\bar{X}}}(\dl+\beta i\bm{J})-\beta\,iX\right)\psi\nn\\
&=& \psi+i \,\hat{\mathcal{O}}_{\bm{J}}(\bm{\bar{X}},X)\,\psi
\eeqa
where $\beta$ is an arbitrary constant and in the last line we have defined the pre-quantum operator $\hat{\mathcal{O}}_{\bm{J}}(\bm{\bar{X}},X)=-i\,\bm{\iota_{\bar{X}}}(\dl+\beta\, i\bm{J})-\beta\,X$. In the case of a degenerate pre-symplectic form, the pairing between $\bm{\bar{X}}$ and $X$ is not unique, so as we have indicated the pre-quantum operator depends on both the function and its associated vector field. The key property of the pre-quantum operator is that the operator commutator algebra is a representation of the associated Lie and Poisson algebras:
\beq
\left[\hat{\mathcal{O}}_{\bm{J}}(\bm{\bar{X}},X)\,,\,\hat{\mathcal{O}}_{\bm{J}}(\bm{\bar{Y}},Y) \right]=i\,\hat{\mathcal{O}}_{\bm{J}}(\bm{[\bar{X}, \bar{Y}]},\{X, Y\})\,.
\eeq
It is standard to take the constant $\beta=-1$, and we will do so as well so that the pre-quantum operator takes the simple form 
\beq
\hat{\mathcal{O}}_{\bm{J}}(\bm{\bar{X}},X)=-i\,\bm{\iota_{\bar{X}}\mathcal{D}_J}+X\ . \label{PQO1}
\eeq

At this point we have a one--to--one correspondence between the classical pair $(\bm{\bar{X}}, X)$ and the pre-quantum operator $\hat{\mathcal{O}}_{\bm{J}}(\bm{\bar{X}},X)$ that respects the associated Lie and Poisson algebras. However, we do not yet have quantum mechanics. Two crucial steps must be taken to regain standard quantum theory. First, one must introduce an inner product with respect to which the operators are Hermitian. At the pre-quantum level, this can be accomplished rather easily by adding an additional term to the operator that takes into account the Lie-drag of the measure under a diffeomorphism. The more difficult task is to choose a polarization that effectively reduces the functional dependence of the pre-quantum wavefunction by half. The standard procedure is to choose a (locally) n-dimensional integrable subspace of the 2n-dimensional tangent bundle denoted $\mathcal{P}(\Gamma)\subset T\Gamma$ such that for any $\bm{\bar{W}^{(i)}}$ and $\bm{\bar{W}^{(j)}}$ in ${\mathcal{P}(\Gamma)}$, we have $\bm{\Omega(\bar{W}^{(i)},\bar{W}^{(j)})}=0$. This polarization is implemented on the pre-quantum wavefunction as follows:
\beq
\bm{\iota_{\bar{W}^{(i)}}\mathcal{D}_J}\psi=0\,. \label{Polarization}
\eeq
This, restricts the wavefunction to be a functional of only $n$ variables. The presence of the covariant derivative in the above ensures that the implementation of the polarization is $U(1)$ covariant.

In the presence of a degenerate symplectic form there are complications in implementing a consistent polarization. One natural way to proceed would be to reduce to the submanifold $\hat{\Gamma}=\Gamma/\mathcal{G}_{ker}$ where $\mathcal{G}_{ker}$ is the set of gauge orbits of the degenerate directions of the symplectic form. The pull-back, $\bm{\hat{\Omega}}$, of the symplectic form to this submanifold is non-degenerate, and the standard procedure can be carried out. However, in practice it is often difficult to characterize the submanifold $\hat{\Gamma}$. Thus, it would be more useful if one had a procedure whereby one could implement a polarization algebra on the pre-quantum Hilbert space over $\Gamma$ that would project down to a consistent polarization on $\hat{\Gamma}$ simply by construction. Fortunately such a procedure has been outlined in the context of degeneracies of the symplectic form on the constraint surface \cite{AAStillerman:GQConstrained}. The degeneracy in our case does not come from the pull-back of the symplectic form to a constraint submanifold, however, the general procedure still applies in our case. Generally speaking, an arbitrary polarization on $\Gamma$ will not project down to a polarization on $\hat{\Gamma}$ without eliminating essential degrees of freedom in the quantum theory. However, one can define a ``constrained polarization" such that the induced polarized Hilbert space over $\hat{\Gamma}$ is consistent and rich enough to be a physically meaningful quantum arena. Fortunately, the procedure is rather simple: one simply needs to enlarge the polarization algebra to include all the degenerate directions of the symplectic form, denoted $\bm{\bar{Z}^{(i)}}$, in the polarization algebra itself so that $\bm{\bar{Z}^{(i)}}=\sum_j {Z^i}_j \,\bm{\bar{W}^{(j)}}$. If one simply implements the new polarization exactly as in (\ref{Polarization}), the polarization on $\Gamma$ projects to a consistent polarization on $\hat{\Gamma}$.

Thus, we conclude that one can work around the problems stemming from a degenerate pre-symplectic form in the geometric quantization procedure so long as we (i) focus on lifting the vector generators of (pre)symplectomorphisms to quantum operators as opposed to functionals, and (ii) take care to implement a consistent ``constrained polarization algebra" on the unreduced phase space. 

\subsection{Pre-quantum operators in the $\bm{\theta}$-representation}
Thus far we have been discussing the pre-quantization scenario in what we will refer to as the $\bm{J}$-representation, where the symplectic one-form is interpreted as a connection on a complex line bundle over the Hamiltonian phase space. However, in keeping with the general theme of this paper, we now shift focus to what we will refer to as the $\bm{\theta}$-representation. The pre-quantization scheme follows completely analogously to the standard procedure described above. In this case, however, we consider a complex line bundle with the \textit{Lagrangian} phase space, $\Gamma_L$ as base manifold. Thus, a typical pre-quantum wave function in the $\bm{\theta}$-representation is a functional of the Lagrangian variables, so for example in the case of Einstein-Cartan gravity we have $\psi=\psi[\varpi, \vep]$. On this bundle we define a $U(1)$ connection $\bm{\mathcal{D}_{\theta}}=\dl+i\bm{\theta}$. The curvature of this connection has support only on the submanifold, $h^{-1}(\Gamma_H)$, where it is proportional to the (pre)symplectic form: $\bm{\mathcal{D}_{\theta}}\bm{\mathcal{D}_{\theta}}=i\bm{\Omega}$. Given a vector field $\bm{\bar{X}}$ that generates a symplectomorphism and its corresponding boundary functional $X$, the pre-quantum operator in the $\bm{\theta}$-representation takes an analogous form to the $\bm{J}$-representation:
\beqa
\hat{\mathcal{O}}_{\bm{\theta}}(\bm{\bar{X}},X)=-i\,\bm{\iota_{\bar{X}}\mathcal{D}_\theta}+X\,. \label{PQO2}
\eeqa
However, it is now understood that $\bm{\bar{X}}\in T\Gamma_L$ and the operator acts on complex functionals over the Lagrangian phase space. Thus, in terms of construction the difference in the two approaches is rather trivial, however, as we will see the new representation has profound implications in the action of the quantum operators on the Hilbert space.

\subsection{Kinematical versus Dynamical constraints}
In order to underscore the differences in the two quantization procedures it will be useful to introduce the notion of kinematical and dynamical
constraints. We will introduce these concepts by considering the constraints of Einstein-Cartan gravity
corresponding to $Spin(3,1)$ gauge transformations and diffeomorphisms at the pre-quantum level. 

It is well known that the kinematics and dynamics of general relativity are indistinguishable in the standard
Hamiltonian approach to general relativity. This is evident in the Hamiltonian constraint, which plays the dual role of
enforcing 4-dimensional diffeomorphism invariance (the kinematics) and at the same time enforcing the non-trivial
parts of the Einstein-equations (the dynamics). The implication is that whereas one can construct a kinematical
Hilbert space corresponding to states invariant under $3$-dimensional diffeomorphisms (an extremely useful arena for
understanding generic background independent properties of quantum spatial geometry), it is impossible to construct a
kinematical Hilbert space invariant under $4$-dimensional diffeomorphisms in the standard framework without simultaneously solving for the dynamical content of the theory.

To illustrate the difference between kinematical and dynamical
operators in this formalism, we first consider the Gauss constraint which implements $Spin(3,1)$ invariance. The
constraint is
\beq
C_{G}(\lambda)=\frac{1}{k}\int_{\Sigma}-D_\omega \lambda\,\star\,e\,e
\eeq
where $\lambda$ is an element of the $spin(3,1)$ Lie algebra. The Poisson algebra naturally is a representation of the
$so(3,1)$ Lie algebra since $\{C_{G}(\lambda_{1}),C_{G}(\lambda_{2})\}=C_{G}([\lambda_{1},\lambda_{2}])$. The
associated vector field (unique up to addition of a vector in the kernel of $\bm{\Omega}$) is
\beq
\bm{\bar{\lambda}}=\int_{\Sigma}-D_{\omega}\lambda\,\frac{\dl}{\dl\omega}+[\lambda,e]\,\frac{\dl}{\dl e}\,,
\eeq
which can be easily verified to satisfy $\bm{\Omega(\bar{\lambda},\ )}=\dl C_{G}(\lambda)$. The corresponding
pre-quantum operator given by (\ref{PQO1}) is 
\beqa
\hat{\mathcal{O}}(\bm{\bar{\lambda}}, C_{G}(\lambda)) &=&- i\,\bm{\iota_{\bar{\lambda}}}\dl- 
\bm{J(\bar{\lambda})}+C_{G}(\lambda)\nn\\
&=& -i\,\bm{\iota_{\bar{\lambda}}}\dl
\eeqa
where the second line follows from the identity $\bm{J(\bar{\lambda})}=C_{G}(\lambda)$. Thus, the implementation
of the constraint as a pre-quantum operator equation gives
\beqa
\hat{\mathcal{O}}(\bm{\bar{\lambda}},C_{G}(\lambda))\,\psi[e,\omega]=0\ \leftrightarrow \ 
\int_{\Sigma}-D_{\omega}\lambda\frac{\dl\psi}{\dl \omega}+[\lambda,e]\frac{\dl\psi}{\dl e}=0\,,
\eeqa
which simply states that the pre-quantum wavefunction must be invariant under infinitesimal gauge transformations as
expected. We refer to such an operator as kinematical since it simply implements a kinematical gauge symmetry.
We can repeat the same procedure in the $\bm{\theta}$-formalism. The constraint is the same and the natural vector
field is $\bm{\bar{\lambda}}=\int_{M}-D_{\varpi}\lambda \, \frac{\dl}{\dl \varpi}+[\lambda,\vep]\,\frac{\dl}{\dl \vep}$.
The result is
\beqa
\hat{\mathcal{O}}(\bm{\bar{\lambda}}, C_{G}(\lambda)) &=& -i\,\bm{\iota_{\bm{\bar{\lambda}}}}\dl
+\bm{\theta(\bar{\lambda})}+C_{G}(\lambda)\nn\\
&=& -i\,\bm{\iota_{\bm{\bar{\lambda}}}}\dl
\eeqa
where we have used the analogous identity $-\bm{\theta(\bar{\lambda})}=C_{G}(\lambda)$. Again, the operator
constraint $\hat{\mathcal{O}}(\bm{\bar{\lambda}},C_{G}(\lambda))\psi[\vep,\varpi]=0$ simply tells us the that the
wave-function must be gauge invariant. Evidently, the pre-quantum operator corresponding to the Gauss constraint in
the $\bm{\theta}$-formalism is also kinematical. The only difference between the two formalisms at this level is the
domain of the wave-function. It can easily be shown that the constraint corresponding the three-dimensional
diffeomorphisms of the boundary manifold $\partial M=\Sigma$ is also kinematical in both formalisms.

Now we turn to the generator of diffeomorphisms in a timelike direction, $\bar{t}$. The associated constraint is the
total Hamiltonian $C_{tot}(\lambda,t)=C_{G}(\lambda)+C_{H}(t)$ where $C_{H}(t)$ is the Hamiltonian constraint given by
\beq
C_{H}(t)=-\frac{1}{k}\int_{\Sigma}\star\,[t,e]\,R\,.
\eeq 
The components of the associated vector field $\bm{\bar{t}}=\int_{\Sigma}\mathcal{L}_{\bar{t}}\omega\,
\frac{\dl}{\dl\omega}+\mathcal{L}_{\bar{t}}e\,\frac{\dl}{\dl e}$
are (partially) defined by Hamilton's equation $\bm{\Omega(\bar{t},\ )}=\dl C_{tot}(\lambda, t)$. In the
$\bm{J}$-formalism, the associated pre-quantum operator is
\beqa
\hat{\mathcal{O}}(\bm{\bar{t}}, C_{tot}(\lambda,t)) &=&- i\,\bm{\iota_{\bar{t}}}\dl- 
\bm{J(\bar{t})}+C_{tot}(\lambda,t)\nn\\
&=&- i\,\bm{\iota}_{\bm{\bar{t}}}\dl-\int_{\Sigma}\star \mathcal{L}_{\bar{t}}\omega\,e\,e+C_{G}(\lambda)+C_{H}(t)\,.
\eeqa
From (\ref{EE2b}) we derive the identity
$\bm{J(\bar{t})}=\frac{1}{2}C_{H}(t)+C_{G}(\lambda)$. Thus, in total the pre-quantum operator in the $\bm{J}$-formalism
is
\beq
\hat{\mathcal{O}}(\bm{\bar{t}}, C_{tot}(\lambda,t))=-i\,\bm{\iota_{\bar{t}}}\,\dl+\frac{1}{2}C_{H}(t)\,.
\eeq
As expected, this operator is not simply a kinematical operator---it implements the non-trivial dynamics in the quantum
theory. Now let us see what the same operator looks like in the $\bm{\theta}$-formalism. In this case, the canonical
vector field is
$\bm{\bar{t}}=\int_{M}\mathcal{L}_{\bar{t}}\varpi\,\frac{\dl}{\dl\varpi}+\mathcal{L}_{\bar{t}}\vep\,\frac{\dl}{\dl\vep}$
where $\bar{t}$ is a vector field on the four-manifold. Using the identity $C_{tot}(\lambda,
t)=-\bm{\theta}(\bm{\bar{t}})$, where we have identified $t\equiv \vep({\bar{t}})$ and $\lambda\equiv -\varpi(\bar{t})$,
the pre-quantum operator is now
\beqa
\hat{\mathcal{O}}(\bm{\bar{t}},
C_{tot}(\lambda,t))\,\psi&=&- \left(i\,\bm{\iota_{\bar{t}}}\dl+\bm{\theta(\bar{t})}+C_{tot}(\lambda,t)\right)\psi\nn\\
&=& -i\bm{\mathcal{L}_{\bar{t}}}\,\psi\,.
\eeqa
Thus, in constrast to the $\bm{J}$-formalism, the operator corresponding to diffeomorphisms in a timelike direction in
the $\bm{\theta}$-formalism is kinematical, similar to the corresponding operator for gauge transformations and
spatial diffeomorphisms. This was to be expected---after all, the wave-function is simply a functional of the
Lagrangian variables, which do not treat space and time on a separate footing. This is in striking contrast to the
standard formalism where timelike diffeomorphism invariance and the physical content of the theory \textit{must} be
solved for simultaneously. More generally, given any Noether vector $\bm{\bar{X}}$ that is also the generator of a symplectomorphism, its associated boundary constraint $C_X=-\bm{\theta(\bar{X})}$ can be lifted to a quantum operator that is kinematical in the $\bm{\theta}$-representation.

It may be argued that, although all of the above results may hold on the pre-quantum Hilbert space, the true
difficulty in geometric quantization comes from the imposition of a polarization of the phase space, which must be
done in order to make contact with standard quantum mechanics, and these results may not hold. However, it is
generally the case that \textit{kinematical} operators in the pre-quantum Hilbert space remain kinematical on the
true, polarized Hilbert space (up to a gauge transformation). To see this, suppose one were to choose a constrained polarization $\mathcal{P}(\Gamma)$. We recall that the defining characteristic of a polarization is that for any $\bm{\bar{W}^{(i,j)}}\in \mathcal{P}(\Gamma)$, we must have $\bm{\Omega(\bar{W}^{(i)}, \bar{W}^{(j)})}=0$. Phrased another way, the pull-back of the symplectic form to $\mathcal{P}(\Gamma)$ is zero \cite{AAStillerman:GQConstrained}. We implement the polarization on the pre-quantum wave-function as in (\ref{Polarization}), which we now write in a more convenient form in the $\bm{\theta}$-representation (the analogous statement holds in the $\bm{J}$-representation):
\beq
\bm{\mathcal{L}_{\bar{W}^{(i)}}} \psi=-i\,\bm{\theta(\bar{W}^{(i)})}\,\psi\,.\label{Polarization2}
\eeq
Now, given a canonical vector field $\bm{\bar{X}}$ associated with a kinematical operator on the pre-quantum Hilbert space, we can always divide up the vector into components perpendicular and parallel to the polarization so that $\bm{\bar{X}}=\bm{\bar{X}_{\perp}}+\bm{\bar{X}_{\|}}$, where $\bm{\bar{X}_{\|}}= X^{\|}_i \bm{\bar{W}^{(i)}}$ and $\bm{\bar{X}_{\perp}}= X^{\perp}_j \bm{Y^{(j)}}$, $(\bm{\bar{Y}^{(i)}}, \bm{\bar{W}^{(j)}})$ form a bases spanning the full tangent space, and each $\bm{\bar{Y}^{(i)}}$ is linear independent of the $\bm{\bar{W}^{(j)}}$. Then, since the operator corresponding to $\bm{\bar{X}}$ is assumed to be kinematical, from (\ref{Polarization2}) we have
\beqa
\hat{\mathcal{O}}_{\bm{\theta}}(\bm{\bar{X}},X)\,\psi &=&-i\bm{\mathcal{L}_{\bar{X}}}\psi\nn\\
&=&-i\bm{\mathcal{L}_{\bar{X}_\perp}}\!\psi-\,X^{\|}_i\,\bm{\theta(\bar{W}^{(i)})}\,\psi\ .
\eeqa
Since the pull-back of the connection to the the polarization submanifold is flat, one can always choose a $U(1)$-gauge such that $\bm{\theta(\bar{W}^{(i)})}=0$ for all $\bm{\bar{W}^{(i)}}$. Thus, in this gauge the operator is still kinematical after the polarization is implemented in the sense that 
\beq
\hat{\mathcal{O}}_{\bm{\theta}}(\bm{\bar{X}},X)\,\psi=0\ \  \longrightarrow \ \ \bm{\mathcal{L}_{\bar{X}_\perp}}\!\psi=0\,.
\eeq

\section{Concluding Remarks}
In this work we have outlined the first few steps in an attempt to make the canonical formulation of classical and
quantum gravity more covariant, both in its retention of the full Lorentz group as the local gauge group, and in its
treatment of spatial and temporal diffeomorphisms. We have isolated the problem of covariance as related to the
degeneracy of the canonical symplectic structure, and shown that when this degeneracy is taken into account, the
closure of the constraint algebra can be demonstrated without reducing the local gauge group to a subgroup of the
Lorentz group, or imposing primary constraints on the phase space variables.

In a follow-up work we will continue the program by computing the constraint algebra explicitly. We will see there
that the constraint algebra is a deformation of the de Sitter, anti-de Sitter, or Poincar\'{e} algebra depending on the sign of the cosmological constant, with the deformation being the local gravitational
degrees of freedom contained in the Weyl tensor. This will explain the vectorial nature of the Hamiltonian
constraint---the vector generators are directly related to the translation generators of the (A)dS/Poincar\'{e} group.

Although this formalism may be useful in classical general relativity, the most apparent advantage of the approach is
in the quantum theory. The advantage of the approach may manifest itself through both its retention of the full local 
Lorentz group and it treatment of temporal diffeomorphisms. With regards to the former, we mention that the discreteness
of area and volumes is intimately related to the compactness of the $SU(2)$
gauge group in standard Loop Quantum Gravity. A more covariant treatment must retain the full, non-compact Lorentz
group and treatments where covariance is retained have seriously questioned the universality of Planck scale
discreteness\cite{Alexandrov:AreaSpectrum, Dittrich:Discreteness, FreidelLivineRovelli:2+1}. This approach may yield
new insight into this open problem.

With regards to the treatment of temporal diffeomorphisms in the quantum theory, the possibility of a \textit{kinematical
spacetime diffeomorphism invariant Hilbert space} may have implications for
the problem of time in quantum gravity. We found that any vector field that generates a bulk symmetry of the action (a Noether vector in our language) is a kinematical operator in the $\bm{\theta}$--representation. In particular, this is true for the generator of time evolution, the total Hamiltonian constraint. The existence of two representations for a class of operators with one of them kinematical is reminiscent of the consistent histories approach to quantum mechanics where one identifies separate intrinsic and extrinsic time evolution operators, with the latter being kinematical \cite{IshamSavvidou, Savvidou:Thesis, Savvidou:GR1, Savvidou:GR2}. It is interesting to see a similar scenario emerging out of geometric quantization, though the connection between the two theories is not fully understood at present.

The existence of a spatial diffeomorphism invariant Hilbert space in quantum
gravity provides a platform for understanding generic properties of quantum geometry apart from the complicated
dynamics imposed on the quantum geometry by the quantum Einstein equations. In this respect,
the existence of a spacetime diffeomorphism invariant Hilbert space may result in the problem of \textit{time} being no
more or less insidious than the problem of \textit{space} in quantum gravity, a problem generally believed to be
synonymous with the
difficulty of finding a classical, continuum limit to quantum geometry. Although the latter problem has proven to be
highly non-trivial, it is a beast of an entirely different nature.

\section*{Acknowledgments}
I would like to thank Martin Bojowald, Abhay Ashtekar, John Engle, and Stephon Alexander for comments, suggestions, and discussions related to this work. This research was supported by NSF grant PHY--0456913 and the Eberly College of Science through the Institute
for Gravitation and the Cosmos at Penn State.
\bibliography{CanLagV2Bib}

\end{document}